\documentclass[prl,aps,floatfix,amsmath,amssymb,superscriptaddress,tightenlines,twocolumn]{revtex4}
\usepackage{graphicx}% Include figure files
\usepackage{epstopdf}
\newcommand{\be}{\begin{equation}}
\newcommand{\ee}{\end{equation}}
\usepackage{amsmath}
\usepackage{amsfonts}
\usepackage{bm}
\usepackage{color}
\usepackage{subfigure}
\usepackage{amsthm}
\usepackage[latin1]{inputenc}

\newcommand{\bra}[1]{\left\langle #1 \right|}
\newcommand{\ket}[1]{\left|#1\right\rangle}

\begin{document}
\title{Long-range entanglement is necessary for a topological storage of quantum information}
\author{Isaac H. Kim}
\affiliation{Institute of Quantum Information and Matter, California Institute of Technology, Pasadena CA 91125, USA}

\date{\today}
\begin{abstract}
A general inequality between entanglement entropy and a number of topologically ordered states is derived, even without using the properties of the parent Hamiltonian or the formalism of topological quantum field theory. Given a quantum state $\ket{\psi}$, we obtain an upper bound on the number of distinct states that are locally indistinguishable from $\ket{\psi}$. The upper bound is determined only by the entanglement entropy of some local subsystems. As an example, we show that $\log N \leq 2\gamma$ for a large class of topologically ordered systems on a torus, where $N$ is the number of topologically protected states and $\gamma$ is the constant subcorrection term of the entanglement entropy. We discuss applications to quantum many-body systems that do not have any low-energy topological quantum field theory description, as well as tradeoff bounds for general quantum error correcting codes.
\end{abstract}

\maketitle

Entanglement entropy is a canonical measure for quantifying entanglement in a bipartite pure state.\cite{Horodecki2009} There has been a recent surge of interest in studying entanglement entropy in a ground state of quantum many-body systems.\cite{Eisert2008} One of the motivations behind these studies is that the entanglement entropy is a useful probe for detecting the phase of the quantum many-body system. For example, entanglement entropy in one dimensional critical systems follows a universal logarithmic scaling law, and its prefactor is related to the conformal charge of the theory.\cite{Calabrese2004} In two spatial dimensions, the quantum dimension of the topological quantum field theory describing the low-energy physics can be inferred from a constant subcorrection term of the entanglement entropy.\cite{Hamma2005,Kitaev2006,Levin2006}

Another important motivation comes from the numerical simulation of quantum many-body systems. Classes of variational ansatz such as the matrix product states,\cite{Fannes1992,Perez-Garcia2007a} projected entangled pair states,\cite{Verstraete2007} and the multi-scale entanglement renormalization ansatz\cite{Vidal2008} have certain entropy scaling laws. Since these variational states reproduce the entanglement scaling of gapped/critical systems, they are suitable for efficiently simulating the ground state properties of the quantum many-body systems. 

Here we show that a certain linear combination of entanglement entropy provides a natural upper bound to the amount of quantum information that can be stored reliably, yet again confirming the importance of this concept. The specific question we are addressing in this Letter is the following: given a set of quantum states $\{ \ket{\psi_i}\}_{i=1, \cdots, N}$ and a physical noise model, how much quantum information can we store reliably? As it stands, the preceding question is too general to give any meaningful answers. However, as soon as one imposes a constraint on the noise model, a lot can be said. The noise model is often {\it local} in nature, meaning that it is described by a sequence of local operations. In order to protect the information against such noises, one must encode information nonlocally: none of the states can be distinguished or mapped into one another via any local operation.\cite{Bravyi2006} These states are called as {\it topologically ordered states}.

There has been a number of results in recent years, where a fundamental limit on the number of topologically ordered states is obtained under a rather general setting.\cite{Bravyi2008,Bravyi2009,Haah2010a,Landon-Cardinal2012,Delfosse2013} The key assumption in these results concerns a property of its parent Hamiltonian: that it can be described by a sum of local commuting Hamiltonians that are frustration-free. One may alternatively assume that the low-energy physics of the quantum many-body system is described by the topological quantum field theory, in which case the number of topologically ordered states is roughly bounded by the number of the emergent quasi-particle types. These two approaches are complementary to each other, but the problem remains completely open if neither of the assumptions hold.

Our result precisely fills this gap: we show that a certain linear combination of entanglement entropy provides a natural upper bound, even without invoking any of the aforementioned assumptions. This approach is able to reproduce some of the known results, but more importantly, it reveals a fundamental connection between the entanglement entropy scaling law and an amount of information encoded in a quantum many-body system.

For concreteness, we state two main assumptions of our approach. First, following Ref.\onlinecite{Bravyi2006}, we assume that there are $N$ states $\{\ket{\psi_i} \}_{i=1, \cdots, N}$ satisfying the topological quantum order(TQO) condition. Formally the set of states satisfies a TQO condition with $(r,\epsilon)$-error  if
\begin{align}
|&\bra{\psi_i} \phi \ket{\psi_j}| \leq \| \phi \|\epsilon, \quad i\neq j \nonumber \\
&|\bra{\psi_i} \phi \ket{\psi_i} - \bra{\psi_j} \phi \ket{\psi_j}| \leq  \epsilon \| \phi \|, \label{eq:TQO}
\end{align}
holds for any operator $\phi$ that is restricted to a ball of radius $r$, where $\| \cdots\|$ is the operator norm. Typically $\{\ket{\psi_i}\}$ is a set of degenerate ground states of a topologically ordered system, and $\epsilon$ converges to $0$ in the thermodynamic limit. If the approximation parameters $r$ and $\epsilon$ are obvious from the context, we shall simply say that the states are locally indistinguishable. An important consequence of Eq.\ref{eq:TQO} is that the reduced density matrices of the locally indistinguishable states are close to each other in a trace distance. Therefore, one can unambiguously define the entanglement entropy of the aforementioned set of states up to a small error, so long as the subsystem can be contained in a ball of radius $r$. We shall call such subsystems to be {\it local}.\footnote{We note in passing that classical error correcting codes in general do not satisfy Eq.\ref{eq:TQO}, {\it e.g.}, a classical repetition code. Therefore, the same argument cannot be applied to such codes.}

The second assumption concerns the property of the entanglement entropy. Recently Grover {\it et al.} argued on a physical ground that the entanglement entropy of a gapped phase should have the following scaling law\cite{Grover2011}:
\begin{equation}
S(A)= a_1 l^{D-1} +a_2 l^{D-2} + \cdots, \label{eq:EE}
\end{equation}
where $A$ is a sufficiently large region with a linear length of $l$, and $D$ is the spatial dimension of the lattice. We shall assume that Eq.\ref{eq:EE} holds for any sufficiently large subsystems. Further, we shall also assume that the leading term of the expansion is proportional to the size of its area. An important consequence of this assumption is that one can cancel out the leading term by making a judicious linear combination of entanglement entropies, see Ref.\cite{Kitaev2006,Levin2006}.

We emphasize that we do not assume anything about the subcorrection terms, aside from their generic scaling laws. Therefore, we are in principle allowing the subcorrection terms to depend on the shape as well as the size of the subsystems. A more precise knowledge about the subcorrection terms can be often obtained, in which case stronger statements can be established. We shall revisit these cases later.

{\it -- A two-dimensional example}

For concreteness, we start with a generic two-dimensional system defined on a torus. As stated before, we assume (i) there are $N$ states that are locally indistinguishable and (ii) the entanglement entropy satisfies Eq.\ref{eq:EE}. For such systems, we obtain the following bound:
\begin{equation}
\log_2 N \leq O(1).\label{eq:2Dbound}
\end{equation}
In other words, under Eq.\ref{eq:EE}, one can only have a constant number of topologically ordered states.

\begin{figure}
\includegraphics[width=3.3in]{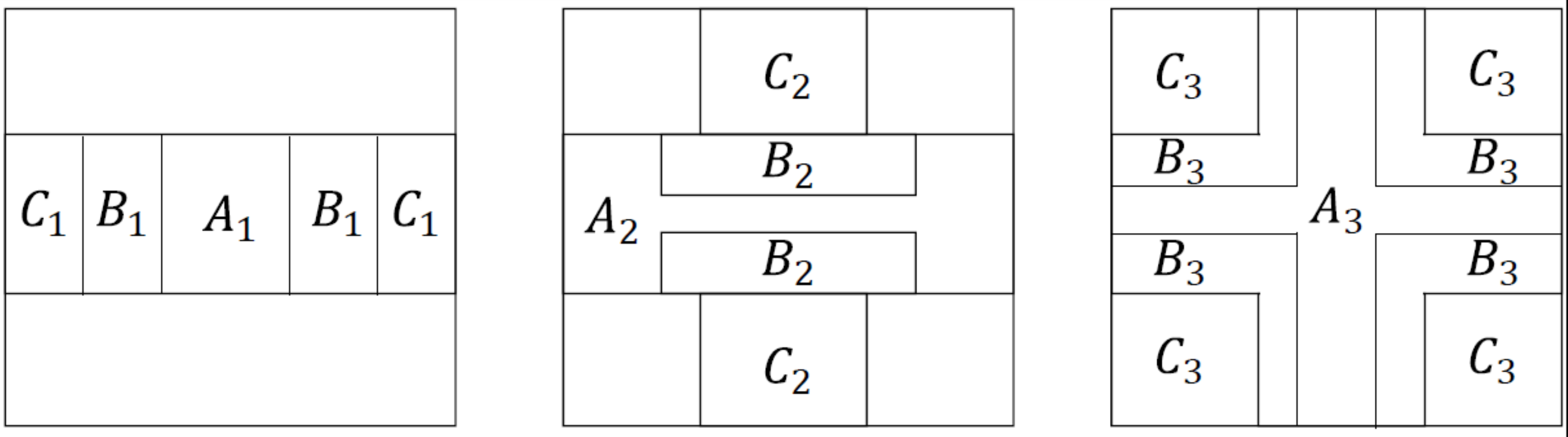}
\caption{Each of the diagrams represent the subsystems $A_i,B_i,C_i$, $i=1,2,3$ with the constraint that $A_iB_iC_i$ is equal to  $A_{i+1}B_{i+1}$. Due to this construction, the entanglement entropy of $A_3B_3C_3$ is the only nonlocal contribution to the sum $\sum_{i=1}^{3} I(A_i:C_i|B_i)$. The entanglement entropy of $A_3B_3C_3$ becomes $\log_2 N$ for a maximally mixed state over $N$ locally indistinguishable states. The rest of the contributions can be computed from the formula for the entanglement entropy of local subsystems, {\it i.e.}, Eq.\ref{eq:EE}.\label{fig:idea}}
\end{figure}

The idea for proving Eq.\ref{eq:2Dbound} is to apply the Markov entropy decomposition(MED) to a maximally mixed state over the $N$ states.\cite{Poulin2010a} More precisely, consider a sequence of subsystems $A_i,B_i,C_i$, $i=1, \cdots , n$ such that (i) $A_i B_i C_i = A_{i+1} B_{i+1}$, (ii) $A_1B_1$ and $B_iC_i$ are {\it local}, and (iii) $A_nB_n C_n$ is the entire system. For such choice of subsystems, the following linear combination of entanglement entropy is nonnegative:
\begin{align}
\sum_{i=1}^{n}I(A_i:C_i|B_i) &=S(A_1B_1)  +  \sum_{i=1}^{n} S(B_iC_i) - S(B_i) \nonumber \\
&-S(A_nB_nC_n). \nonumber
\end{align}
Here $I(A:C|B) := S(AB) + S(BC) - S(B) - S(ABC)$ is the quantum conditional mutual information, which is known to be always nonnegative due to the strong subadditivity of entropy(SSA).\cite{Lieb1972} By choosing the global state to be a uniform mixture of the $N$ locally indistinguishable states, {\it i.e.}, $\sum_{i=1}^{N} \frac{1}{N} \ket{\psi_i}\bra{\psi_i}$, we arrive at the following bound:
\begin{equation}
\log_2 N \leq S(A_1B_1) + \sum_{i=1}^{n} S(B_iC_i) - S(B_i). \label{eq:SSA_bound}
\end{equation}
Since $A_1B_1$, $B_iC_i$, and $B_i$ are all local, their entanglement entropy can be replaced by an entanglement entropy of one of the states $\ket{\psi_j}$ with a small correction. The correction term can be estimated by using Fannes inequality\cite{Fannes1973} which holds for any quantum states $\rho, \sigma$ supported on a $d$-dimensional Hilbert space:
\begin{equation}
|S(\rho) - S(\sigma)| \leq \epsilon \log_2 d - \epsilon \log_2 \epsilon, \quad \epsilon :=|\rho - \sigma|_1, \label{eq:FannesInequality}
\end{equation}
where $|\cdots |_1$ is the trace norm. Eq.\ref{eq:2Dbound} can be derived by choosing an appropriate set of subsystems such that the boundary contributions  cancel out, while the remaining terms may survive. One choice of such subsystems is depicted in FIG.\ref{fig:idea}.
Derivation of Eq.\ref{eq:2Dbound} clearly follows from Eq.\ref{eq:EE} and Eq.\ref{eq:SSA_bound}. The leading terms of the entanglement entropy cancel out each other. The remaining terms contribute a constant amount for each subsystems. Since a constant number of subsystems was involved in the right hand side of Eq.\ref{eq:SSA_bound}, the contributions altogether can be bounded by some constant.\footnote{More precisely, the number of subsystems is a constant that is independent of the system size.}

{\it -- Higher dimensional systems}

It should be clear from the analysis of the two-dimensional systems that the same argument can be applied to higher dimensional systems as well. That is, the leading term of the entanglement entropy cancels out due to the choice of the subsystems made in the right hand side of Eq.\ref{eq:EE}. Since there are constant number of subsystems involved in this construction, and the leading surviving terms grow as $O(L^{D-2})$, we arrive at the following bound:
\begin{equation}
\log_2 N \leq O(L^{D-2}). \label{eq:TEE_D}
\end{equation}
This bound is unlikely to be saturated by any models that are described by BF theory.\cite{Baez2000} These models have a constant number of topologically ordered degenerate ground states, whereas the right hand side of Eq.\ref{eq:TEE_D} grows with the system size.

However, we emphasize that there are exotic topologically ordered systems in three dimensions that do saturate Eq.\ref{eq:TEE_D}, up to a multiplicative constant. Interesting examples include Chamon's model and Haah's cubic code, which are known to have ground state degeneracies that increase as $N=2^{cL}$ for certain choices of $L$, where $c>0$ is some constant.\cite{Chamon2005,Bravyi2010b,Haah2011} This result has at least two implications. First, the entanglement entropies of such models must have an extensive subcorrection term that is linear in $L$. Second, these linear contributions cannot be canceled out by each other: had that been the case, the right hand side of Eq.\ref{eq:TEE_D} would be $O(1)$, which violates the $O(L)$ lower bound. In other words, there is a subleading contribution to the entanglement entropy that cannot be canceled out in any ways from any choice of subsystems. The physical meaning and the origin of these contributions is unclear at this moment.

{\it -- Bounds for general quantum error correcting codes}

In general, one cannot expect the leading terms of the entanglement entropy to be canceled out by choosing an appropriate set of subsystems, especially for critical systems and ground states of nonlocal Hamiltonian. We show that, even for such generic systems, a nontrivial tradeoff bound can be obtained.

There are usually three standard parameters in addressing such tradeoff bounds. The first parameter is $n$, which is the number of particles. Implicit in this assumption is that the state of each particle can be described by a Hilbert space with a bounded dimension, say $2$.  The second parameter is $d$, which denotes the {\it code distance} of the quantum error correcting code. Code distance is a minimal number of local operation that can map one of the states of the code to another (orthogonal) one. For example, Kitaev's two-dimensional toric code\cite{Kitaev2003} on a $L \times L$ grid has a code distance of $L$: in order to map one of the ground states to another one, one must create a particle-antiparticle pair, and fuse them together to form a nontrivial loop. Since any local operation can transport a particle by a constant amount of distance, the minimal number of local operation clearly scales with $L$. The last parameter is $k=\log_2 N$, the {\it number of encoded qubits}. Needless to say, $k$ quantifies the amount of information that can be stored in the code.

A notable bound was derived by Bravyi {\it et al.}\cite{Bravyi2009}:
\begin{equation}
kd^{\frac{2}{D-1}} \leq O(n), \label{eq:CLH_bound}
\end{equation}
under the assumption that the quantum error correcting code is spanned by a set of degenerate ground states of a Hamiltonian with a special structure.\footnote{More specifically, they assumed that the Hamiltonian can be expressed as a sum of local commuting projectors.}

Here we derive the following alternative bound:
\begin{equation}
kd^{1-\alpha} \leq O(n), \label{eq:subvolumelaw_bound}
\end{equation}
under the assumption that entanglement entropy satisfies a subvolume law:
\begin{equation}
S(A) = O(|A|^{\alpha}), \quad 0 \leq \alpha \leq 1. \label{eq:subvolume}
\end{equation}
Eq.\ref{eq:subvolumelaw_bound} is a simple consequence of Eq.\ref{eq:SSA_bound} and the subadditivity of entropy, {\it i.e.}, $I(A:B) = S(A) + S(B) - S(AB) \geq 0$. These two bounds together yield the following inequality:
\begin{equation}
k \leq \sum_i S(X_i),  \quad |X_i| <d   \label{eq:rate_entropy_inequality}
\end{equation}
for a quantum code with a code distance $d$ and a number of encoded qubits $k$, where $\{X_i \}$ is a certain partition of the system. Setting each of the subsystem sizes to be $O(d)$, and applying Eq.\ref{eq:subvolume} to Eq.\ref{eq:rate_entropy_inequality}, one can derive Eq.\ref{eq:subvolumelaw_bound}. There are at least two important differences between Eq.\ref{eq:CLH_bound} and Eq.\ref{eq:subvolumelaw_bound}. On one hand, Eq.\ref{eq:subvolumelaw_bound} is more general than Eq.\ref{eq:CLH_bound} in that it does not require any structure about the parent Hamiltonian. On the other hand, Eq.\ref{eq:subvolumelaw_bound} provides a weaker tradeoff bound than Eq.\ref{eq:CLH_bound} does.

Dividing both sides of the inequality by $n$, we find that the {\it rate} of a quantum error correcting code $\frac{k}{n}$ is bounded by the average entanglement entropy per volume over any partitions $\{X_i \}$, $|X_i| <d$. Our result shows that studying the entanglement properties of a quantum error correcting code is a relevant problem for understanding its fundamental limit. In particular, Eq.\ref{eq:rate_entropy_inequality} gives a necessary condition for a quantum error correcting code to have a nonvanishing rate: its average entanglement entropy over subsystems smaller than the code distance must satisfy a strict volume law.

{\it -- Benchmark with the known results}

Here we reproduce some of the known results in the literature in order to evaluate the strength of our approach.  First, consider a two-dimensional gapped system on a torus that is described by a topological quantum field theory. Slightly away from the fixed point, the entanglement entropy satisfies the following relation,
\begin{equation}
S(A) = al - b_0(A)\gamma +O(\frac{1}{l}), \label{eq:strict_arealaw}
\end{equation}
where $l$ is the boundary area of $A$, and $\gamma$ is the topological entanglement entropy.\cite{Kitaev2006,Levin2006} $b_0(A)$ is the number of connected of components of the boundary of $A$. 

In the previous analysis, we did not assume any {\it a priori} knowledge about the subcorrection term. Instead, by directly applying Eq.\ref{eq:strict_arealaw} to Eq.\ref{eq:SSA_bound}, we arrive at the following bound:
\begin{equation}
\log_2 N \leq 2\gamma + O(\frac{1}{l}) \label{eq:TEE_onesided_stability},
\end{equation}
where $l$ is the linear size of each subsystems. Recall that the topological entanglement entropy is related to the total quantum dimension of the system:
\begin{equation}
\gamma = \log_2 \sqrt{\sum_a d_a^2}, \nonumber
\end{equation}
where $d_a$ is the quantum dimension of a particle with a topological charge $a$.\cite{Kitaev2006,Levin2006} Eq.\ref{eq:TEE_onesided_stability} confirms the intuition that an amount of long-range entanglement limits the topological ground state degeneracy.

{\it -- Outlook}

MED provides a powerful numerical framework for studying finite-temperature quantum many-body systems.\cite{Poulin2010a,Ferris2012} By extending the observation of Hastings and Poulin, we have shown that MED can be also used elucidate a connection between entanglement entropy and a number of topologically ordered states. In particular, we have shown that there is a fundamental reason behind the existence of the topological entanglement entropy, even without using the formalism of topological quantum field theory. After all, a state with a sufficiently small topological entanglement entropy cannot have another state that is locally indistinguishable from the original state. It is quite remarkable that such a tight tradeoff bound can be obtained simply from the local property of the wavefunction alone. There are several interesting physical systems for which our approach might be useful.

For example, our approach should be easily generalizable to finite temperature systems. In such systems, the topological entanglement entropy is interpreted as an order parameter characterizing the phase.\cite{Castelnovo2007,Castelnovo2008,Iblisdir2010,Mazac2011} While the technical details must be worked out, we can have an intuitive explanation on why these terms arise, even without resorting to the standard topological quantum field theory argument. Some of these models become a self-correcting classical/quantum memory depending on the temperature.\cite{Alicki2008,Chesi2009} Since the ``logical operators" that are associated to these systems are highly nonlocal, two distinct states cannot be distinguished via any local operation. If the topological entanglement entropy becomes sufficiently small, the number of locally indistinguishable states is bounded by a small number, which would give rise to a contradiction. It is well-known that Eq.\ref{eq:EE} is modified if the subsystem contains a quasi-particle with a nontrivial topological charge.\cite{Kitaev2006} It would be interesting to extend our analysis to such settings as well.

We conclude with a remark that our result can be used as a rigorous tool for proving a distinctiveness of different quantum many-body phases. Given a quantum many-body system on a torus with $N$ topologically protected ground states, it cannot be adiabatically connected to {\it any} state with a topological entanglement entropy strictly lower than $\frac{\log_2 N}{2}$. This is due to the fact that topologically protected states remain to be so under an adiabatic evolution.\cite{Bravyi2006,Bravyi2010} Therefore, the lower bound for the topological entanglement entropy, {\it i.e.}, Eq.\ref{eq:TEE_onesided_stability}, remains stable with a small correction that vanishes in the thermodynamic limit.

{\it -- Acknowledgements}

I would like to thank John Preskill, Alioscia Hamma, and Roger Mong for helpful discussions and suggestions.  Part of this work was done while the author was attending 2013 Coogee quantum information theory workshop. This research was supported in part by NSF under Grant No. PHY-0803371, by ARO Grant No. W911NF-09-1-0442, and DOE Grant No. DE-FG03-92-ER40701.

\bibliography{bib}

\end{document}